\documentclass[12pt]{elsarticle}
\usepackage{graphicx,color} 
\usepackage{natbib}

\usepackage{amsmath,amsfonts,amssymb,theorem}
\usepackage{url,subfig,anysize,wrapfig}
\usepackage{multirow,tabularx,afterpage,fancyhdr}
\usepackage{cases}
\usepackage{epstopdf}
\usepackage{bm}
\usepackage[hmargin=1in,vmargin=1in]{geometry}

\usepackage[linewidth=1pt]{mdframed}
\usepackage{lipsum}
\usepackage{lineno}
\usepackage{placeins}

\usepackage{tikz}
\usetikzlibrary{calc}

\usepackage{cleveref} 
\usepackage{mathrsfs}
\usepackage[titletoc]{appendix}
\usepackage{verbatim} 
\usepackage{amssymb}
\usepackage{lineno}

\newcommand{\degree}{^\circ}

\journal{}

\begin{document}

\begin{frontmatter}

\title{A method to measure the embedded crack length and position in high-density polyethylene using microseconds ultrasound time signal}

\author[main]{Sijun Niu}
\author[main,main2]{Venkatsai Bellala}
\author[main]{Daanish Aleem Qureshi}
\author[main,main2]{Vikas Srivastava\corref{cor1}}
\ead{vikas\_srivastava@brown.edu}
\cortext[cor1]{Corresponding author}
\address[main]{School of Engineering, Brown University, 184 Hope Street, Providence, RI 02912, USA}
\address[main2]{Center for Biomedical Engineering, Brown University, 184 Hope Street, Providence, RI 02912, USA}

\begin{abstract}
High-density polyethylene (HDPE) is a semi-crystalline polymer used in several critical applications, ranging from cooling water pipelines in nuclear power plants and distribution pipelines for natural gas and hydrogen to biomedical implants. Embedded crack-like flaws form within HDPE during fabrication or operations. Non-visible flaws grow over time and can cause catastrophic failure if undetected. Large structures such as HDPE pipelines where the location of a flaw is not known require a fast, non-destructive evaluation (NDE) method where the sensor can move rapidly across the structure with very short microseconds at each location. This is only possible if the flaw is evaluated in HDPE and other polymeric structures using microseconds of time signal. Ultrasonic A-scan (time signal) allows for the rapid scan of large structures. 
We propose and show the accuracy of a machine learning-based Ultrasound NDE method that can rapidly and accurately predict embedded crack length and position simultaneously in HDPE with only tens of microseconds of time signal sensing. Current NDE methods rely on technical experts to evaluate the ultrasound measurements, which leads to high uncertainty and errors as the quantitative information about a crack is subtly encoded in the reflected signal. A method to quantify crack size in HDPE and other polymers using a very short Ultrasound time signal is lacking. We suggest that an optimally trained machine learning model can decipher the crack characteristics using short measures of time signal, but a lack of large, well-distributed, and labeled datasets to train machine learning models continues to be a major limitation. To overcome this limitation, we have conducted computer simulations of ultrasound on HDPE to develop training data. We show that fully finite element simulations trained convolutional neural network (CNN) can accurately predict crack lengths and positions in HDPE from experimentally measured ultrasound A-scan microsecond signals, with an average error of 3.2\% for the crack lengths and 3.8\% for the crack positions. Our method is based on the 1D time amplitude signal acquired over a very short time period and not based on 2D image analysis as the image rendering NDT is very slow and susceptible to losing subtle but important crack feature information during the post-processing to create images. The proposed methodology presents a pathway for training CNN using computationally generated data and applying the trained CNN in the field to quantify hidden cracks in large HDPE or other polymer structures using ultrasound time signals when the measurement window is very small. 
\end{abstract}


\end{frontmatter}



\section{Introduction}
Polyethylene (PE) is a semi-crystalline polymer with repeating ethylene monomer units in the polymer chain. PE consists of a highly ordered and molecularly structured crystalline phase and an amorphous phase of entangled disordered chains. High-density polyethylene (HDPE) is a highly linear variant of PE as it lacks branches, enabling close packing of its polymer chains. Similar to PE, it is a low-cost thermoplastic polymer consisting of a mixed amorphous and crystalline structure and has high crystallinity, strength, and moderate stiffness. HDPE is used in various biomedical and industrial applications \cite{Demirors2011ThePolyethylene, Paxton2019BiomedicalPolyethylene}. Within the medical device space, HDPE is used for bone grafts, surgical implants, and total joint replacements \cite{Paxton2019BiomedicalPolyethylene,Fouad2013Thermo-mechanicalAgeing}. HDPE pipes are used for road drains  \cite{Stuart2011EvaluationConstruction, Goddard2018GrowthApplications}, and natural gas distribution pipes \cite{Gong2021, Tutunchi2020, Taous2020} due to their resistance to cracking and chemical interactions and because of their low cost and ease of installation \cite{Demirors2011ThePolyethylene}. HDPE has been utilized to transport cooling water in nuclear power plants \cite{Zheng2018APlant}. HDPE, due to its chemical resistance to hydrogen, will play a crucial role in clean energy hydrogen storage and transport \cite{Iskov2010FieldGrid}. 

HDPE is easily recyclable \cite{Nguyen2021Long-termReview}, and has long-term durability and performance  \cite{Lang1997ApplicabilityPressure, Kausch1983TheThermoplastics,Nguyen2021Long-termReview}. However, flaws introduced in manufacturing \cite{Gassman2005FieldPipe} and during operations can weaken the polymer structure, reducing its performance and life cycle. Extrusion defects were found to be responsible for the majority of crack initiations in HDPE tubes \cite{Schouwenaars2007SlowHDPE-tubes}. HDPE pipes in highway drainage systems showed significant internal cracks during operations \cite{Gassman2005FieldPipe}, and HDPE pipe joints, such as those formed using butt-fusion, are prone to embedded defects \cite{ShafieiAlavijeh2021UsingPipes}. Even under low stresses, HDPE is susceptible to slow crack growth \cite{SCHOUWENAARS2007}. Detection and characterization of cracks or crack-like flaws become essential to assess the long-term integrity of the HDPE structure. 
Crack length is the most important characteristic determining the load or stress under which an HDPE structure will fail. For a crack of length $a$, the crack tip opening displacement rate $\dot{\delta}$ for a slowly growing crack in PE can be estimated using equation 1 \cite{Brown1995APolyethylene}. Chan and Williams \cite{Chan1983SlowPolyethylenes} experimentally demonstrated that linear elastic fracture mechanics can approximately describe the failure stress $\sigma_f$ in PE, which indicates the relation shown in equation 2.

\begin{align}
    \dot{\delta} &\propto a^m \qquad \text{where}\; m>0\qquad\text{(Slow Crack Growth Model)}\\
    \sigma_f &\propto a^m\qquad\text{where}\; m=-\dfrac{1}{2}\qquad\text{(Linear Elastic Fracture Mechanics)}
\end{align}

A variety of tests can be applied to assess the long-term health and performance of HDPE structures. Environmental stress crack resistance and time-temperature superposition test methods can estimate the chemical and mechanical/thermal degradation profile of HDPE specimens. However, these test methods are destructive and not very useful for in-field measurements \cite{Troughton2010APipes}. Non-destructive testing (NDT) or non-destructive evaluation (NDE) allows the detection of previously unknown cracks and flaws without damaging the specimen. Visual and other surface examination techniques can identify surface damages \cite{Zheng2018APlant} but are ineffective for embedded flaws. Infrared thermography tests for flaws in HDPE pipes and tanks by analyzing heat dissipation from damaged regions, but its accuracy is limited \cite{Behravan2023FieldMethods}. Microwave imaging has been shown to be capable of detecting internal flaws and voids in HDPE pipes \cite{Crawford2010PreliminaryTesting, Stakenborghs2009MicrowaveWelds} but is limited by low resolution. Ultrasound is a promising NDE method. Ultrasound methods are frequently utilized to image and detect embedded cracks in HDPE \cite{Stakenborghs2009MicrowaveWelds,Yao2014CrackOverview,Postma2012SuitabilityJoints,Zhu2015Non-DestructiveUltrasound, Frederick2010High-DensityArray}. An ultrasound device propagates waves through a specimen, and waves that reflect from an internal anomaly are analyzed to discern the anomaly. A common ultrasonic flaw detection approach is the pulse-echo technique: reflections of the generated wave from flaws and cracks are recorded as echoes or peaks in a scan. An A-scan is a simple time–amplitude ultrasound signal. B-scan, on the other hand, produces a 2D cross-sectional image of the anomaly from the reflected ultrasound signals. A B-scan requires significantly longer data collection time at a location and depends on the post-processing of raw data to create 2D images. \emph{Compared to image based NDE methods, microseconds raw time signal sensing (A-scan) provides fast scanning essential for evaluating large structures. A time signal based NDE method is also not limited by approximation errors (loss of important information contained in very subtle signal variations that account for quantitative measures of the change in the size of the flaw) made by post-processing algorithms. In this paper, we show that the microseconds of the flaw-reflected ultrasound time signal for HDPE contain sufficient information to accurately quantify the length and position of an embedded non-visible crack.} Machine learning can be applied to interpret ultrasound signals to eliminate technician error \cite{Oishi2001NeuralUltrasonics}. Machine learning has been applied to measure embedded crack characteristics in steel from ultrasound measurements \cite{Niu2022UltrasoundNetwork,Niu2022SimulationMeasurements}. Machine learning is used in fields ranging from computer vision and speech recognition \cite{Liu2017AApplications} to problems in mechanics \cite{Bock2019AMechanics,Niu2023ModelingPerformance,kuhl2023,LinkaKuhl2023,Hsu2020UsingSolids,Wang2023,Chan2022}. Convolutional neural networks (CNNs) have efficient pattern recognition, and image classification, especially when the dimensions of the raw inputs are very large and discerning high-level features is essential \cite{LeCun2015DeepLearning,Zeiler2014VisualizingNetworks}. CNNs have been used in identifying the presence of cracks in butt-fused joints \cite{ShafieiAlavijeh2021UsingPipes} but have been unable to quantify the crack length accurately.

Applying rapidly obtained, unprocessed microseconds of ultrasound time signal to accurately quantify key hidden cracks features in \emph{polymers} has not been done. We present a method that applies fully computational simulations trained CNN to accurately quantify the length and position of penny-shaped embedded cracks in semicrystalline polymer HDPE from reflected ultrasound time signal NDT.  It is not feasible to create well-labeled experimental NDT datasets for embedded non-visible cracks \cite{CANTEROCHINCHILLA2022}. This hampers the application of machine learning for crack quantification in HDPE. To fill this gap, finite element simulations using Abaqus software were developed for HDPE to create training datasets for CNN. A commercial ultrasound test unit was used to obtain A-scan measurements on HDPE specimens containing embedded cracks. The fully computationally trained CNN was applied to the independent experimental signals. The CNN can predict the length and position of embedded elliptical cracks in HDPE samples with good accuracy.

\section{Computational Method}
\label{computational}

\subsection{Finite element simulations for ultrasonic NDT}\label{simulation}

\begin{figure}[h!]
        \centering
        \includegraphics[width=0.5\textwidth]{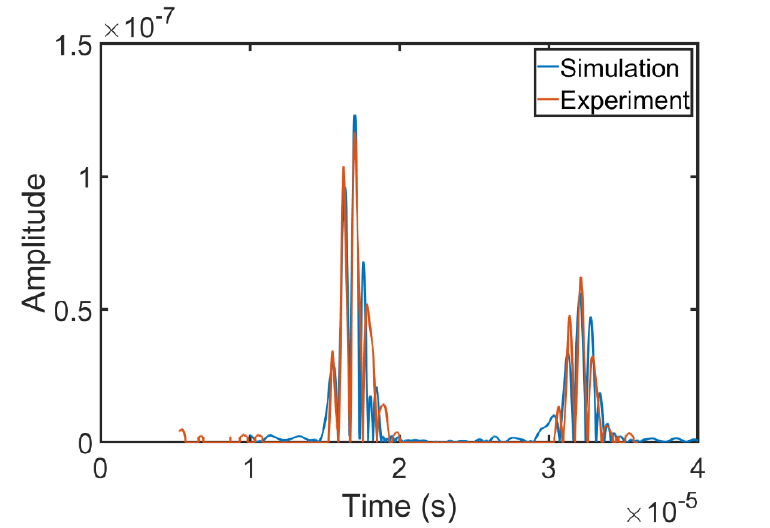}
        \caption{Experimental and simulation ultrasound signals for a 12.7 mm thick flat sheet of HDPE without any cracks.}
        \label{fig:hdpe_flat_signal}
    \end{figure}

Finite element simulations were conducted using Abaqus to first evaluate acoustic attenuation and dispersion of the transmitted A-scan ultrasound signals within HDPE specimens and then create a simulation-based training dataset for CNN. Finite element methods have been successfully applied to various engineering applications and proven effective for yielding accurate and reliable results \cite{KONALE2023, Srivastava2010SMP, Bai2021AEngineering,  Zhong2021AGrowth,  Kothari2019ARange, Srivastava2011STRESSPIPE}. Assumptions of elasticity for long ultrasound wave propagation distances generally do not hold for HDPE as they do for metals such as steel \cite{Niu2022UltrasoundNetwork}. Viscoelastic behavior may affect ultrasound signals due to dissipation. By measuring the acoustic attenuation of A-scan ultrasound signals, we can evaluate the extent to which viscoelasticity impacts the characterization of crack length. Initial simulations were conducted for a 12.7 mm thick flat sheet with no crack. HDPE's material properties were chosen with assumptions of linear elasticity. Young's modulus was 0.97 GPa,  Poisson's ratio was 0.43, and the density was 954 kg/m$^3$. These material properties are typical for HDPE  \cite{Brown1995APolyethylene}. A calibration experiment was also performed on a 12.7 mm thick flat sheet of HDPE using a 1 MHz transducer on Olympus Epoch 650 Ultrasound NDT equipment (Figure \ref{fig:ultrasound_equip}). For Natural gas, water and many other pipeline applications, 12.7 mm represents a commonly used pipe thickness. Figure \ref{fig:hdpe_flat_signal} compares the ultrasound signals obtained from the simulation and the calibration experiment. The two peaks represent the first and second back-wall echos. Note that the initial pulse was omitted here, as its amplitude is outside of the ultrasound equipment's data acquisition limit.

Signal attenuation was determined to be negligible for the 1 MHz frequency and 25.4 mm and 50.8 mm acoustic travel distance in this study. This is evidenced by similar maxima at each peak in both the simulated and experiment ultrasound signals. The width of each echo also indicates insignificant signal dispersion. Therefore, we can conclude that for small acoustic travel distances and frequencies near 1 MHz, signal attenuation and dispersion negligibly affect the final ultrasound signals. The HDPE response can thus be assumed to be dominantly elastic in this regime. This observation is consistent with linear elastic conclusions in the slow crack growth studies of Chan and Williams \cite{Chan1983SlowPolyethylenes}. All simulations for this study assumed linear elastic material properties for HDPE.

\begin{figure}[h!]
        \centering
        \includegraphics[width=0.75\textwidth]{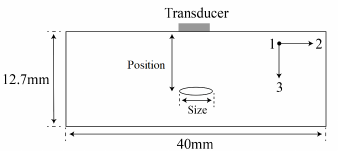}
        \caption{Cross-sectional view illustrating two key crack properties of the HDPE block on its symmetry plane. These properties are the crack length \textit{a} and the crack position \textit{d}.}
        \label{fig:geometry_diagram}
    \end{figure}

We consider the crack length and crack position as two main parameters to quantify for embedded cracks in HDPE. We consider a cuboid (block) geometry mimicking a small section of a large span structure or a section of a large radius-to-thickness ratio pipe. The crack is taken to have a penny-shaped, elliptical geometry. The crack length is defined as the major axis of the crack, and the crack position is defined as the distance between the transducer (measurement surface) and the crack. The crack is taken to be oriented horizontally in the plane that has 12.7 mm thickness. The crack's minor axis was fixed at 0.5 mm, and the crack length varied from 1 to 6 mm. This provides a ratio of crack length to thickness range between 2 and 12. Additionally, the crack position varied between 3 and 11 mm. Evenly dividing the ranges for crack length and crack position into 40 intervals yields a dataset containing $40\times 40 = 1600$ simulated ultrasound signals. Table \ref{tab:crack_range} summarizes the crack property ranges used in our finite element simulations, and Figure \ref{fig:geometry_diagram} shows the geometry. 

    \begin{table}[h!]
        \centering
        \caption{Embedded crack length and position range in HDPE considered in this study.}
        \begin{tabular}{c|c|c}
            \hline
            {Parameter} & {Length} & {Position}\\\hline
            {Min} & 1 mm & 3 mm\\
            {Max} & 6 mm & 11 mm\\
            \hline
        \end{tabular}
        \label{tab:crack_range}
    \end{table}

\begin{figure}[h!]
        \centering
        \includegraphics[width=\textwidth]{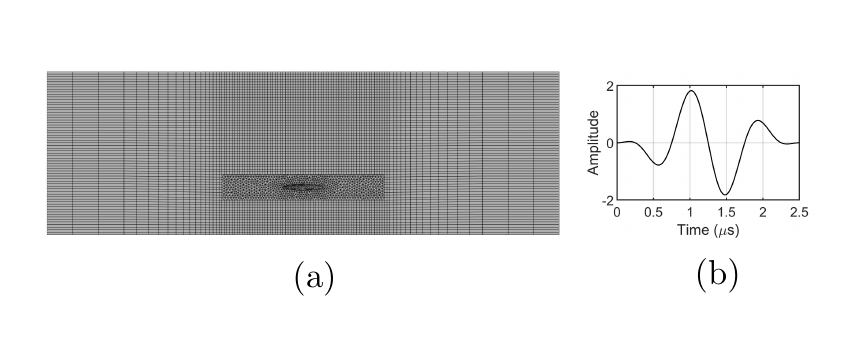}
        \caption{(a) Cross-sectional view showing the finite element mesh of the block on its symmetry plane. The mesh becomes significantly finer as it approaches the central region containing the elliptical flaw. (b) Amplitude versus time plot of the simulated pulse generated by the time-dependent pressure boundary condition at the top surface nodes representing the 12.7 mm diameter circular probe region on the block. }
        \label{fig:finite_mesh}
    \end{figure}

Next, finite element simulations were performed using Abaqus/Explicit to reproduce ultrasound NDT for HDPE on a computer. The HDPE geometry was defined as a $40\times40\times12.7$ mm rectangular cuboid. The ultrasonic signal propagation was directed in the 12.7 mm thickness dimension, with a frequency of 1 MHz and 2.5 mm wavelength. Since this cuboid is symmetrical, only half of it was simulated. The center of the simulated geometry contains an embedded flaw. Because the speed of sound in HDPE is rather fast (2340 m/s), the element sizes need to be sufficiently small to ensure numerical stability. Thus a very fine mesh with an element size of 0.1 mm was used near the flaw, with the element size gradually increasing in the domain far from the flaw. Since the geometry near the flaw is irregular, the central region containing the flaw was meshed using C3D10M tetrahedral elements. The remainder of the geometry utilized C3D8R brick elements to improve computational efficiency away from the flaw. Figure \ref{fig:finite_mesh}(a) depicts the cross-section for a typical mesh used with this geometry. The total number of elements in each finite element simulation was between three and four hundred thousand. The total simulation time was 20 microseconds, and the time step was fixed at 10 nanoseconds. Consistent with the transducer size available for experiments, a time-dependent pressure boundary condition was employed on a 12.7 mm diameter circular region. The ultrasound pulse was simulated with a raised-cosine type waveform at 1 MHz frequency, as shown in figure \ref{fig:finite_mesh}(b). The period of the signal is 2.5 $\mu$s, and the longitudinal wave travels in the 12.7 mm thickness direction. 
    
\subsection{Simulation trained convolutional neural network}\label{cnn}

The CNN is a feed forward neural network with an efficient architecture ideal for applications in imaging, speech and signal processing, and more. CNN and other machine learning techniques are also becoming increasingly useful for structural mechanics problems. Machine learning has been used to solve physics-informed partial differential equations \cite{Raissi2019Physics-informedEquations}, model fluid flow \cite{Brunton2020MachineMechanics}, predicting mechanical properties with continuum mechanics approach \cite{Lu2020ExtractionIndentation,Li2019PredictingLearning}, and ascertaining constitutive properties \cite{Huber2020Editorial:Science}. 2D image-based CNNs have been used to discern properties of surface cracks from ultrasonic images of the cracks \cite{Pyle2021DeepNDE}. CNNs can also process noisy ultrasound signals with high accuracy for the classification of weldment flaw defects \cite{Munir2019ConvolutionalConditions}. Krokos et al. \cite{Krokos2022AFeatures} applied a Bayesian multiscale CNN to reduce computational costs for modeling structures with microscale features. Evaluation of the degree of structural damage was studied using echo state networks, and multi-scale CNN \cite{He2022}. Meng et al. \cite{Meng2017UltrasonicNetworks} have proposed an automatic ultrasonic signal classification system using a deep neural network for defective carbon fiber composite systems. 

A typical CNN architecture consists of three layers: the convolutional layer, the pooling layer, and the fully connected (FC) layer \cite{Wen2018AMethod}. Inputs are passed first to the convolutional layer, where a kernel or multi-channel filter obtains high-level information (feature extraction). The features are down-sampled in the pooling layer in order to reduce dimensions. Learning occurs in the FC layer, where all the neurons are connected. Assigning an activation function to each neuron in the convolutional and FC layers allows for nonlinear learnability from extracted features \cite{Zhang2018ALoad}. The ReLu activation function is one common choice and was selected for this study due to its fast convergence in CNNs, and notably great performance in deep learning applications \cite{Maas2013RectifierModels,Nair2010RectifiedMachines}.

\begin{equation}
    \text{ReLu}(z) = max(0,z)
\end{equation}

\begin{figure}[h!]
    \centering
    \includegraphics[width=\textwidth]{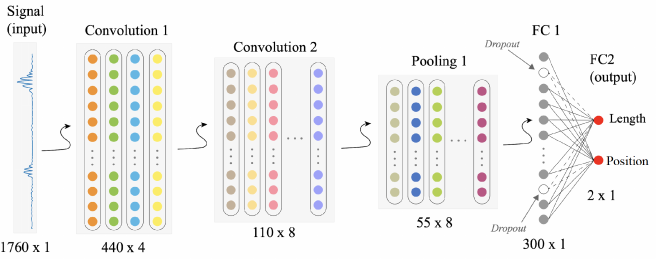}
    \caption{CNN architecture for quantification of embedded crack length and position in HDPE from ultrasound A-scan signals.}
    \label{fig:cnn_layout}
\end{figure}

The type of pooling layer also needs to be selected. The max-pooling layer selects local maxima among input features to speed up training and reduce dimensions. Our architecture utilizes a max-pooling layer. Dropout is a technique used to prevent network overfitting \cite{Srivastava2014}, in which some neurons within a layer have a probability $p$ to be deactivated during training. Dropout is effective in regularizing and providing better generalization ability to the network. We have applied dropout to the FC layer. The CNN architecture used in this study (Figure \ref{fig:cnn_layout}) consists of two convolutional layers, one pooling layer, and two FC layers, including the output layer. We have used ReLu in both convolutional layers and the first FC layer. The optimization algorithm utilized was Adam, and a dropout probability of 0.2 was used in the first FC layer. The mean squared error (MSE) loss function, shown below, was selected as our loss function. Finally, this CNN was trained for 2000 epochs using a learning rate of 0.0005. Table \ref{tab:cnn_config} describes the configuration of our CNN in more detail.

\begin{align*}
    \text{Loss}_\text{MSE} &= \sum_{i=1}^n \left(y_i-t_i\right)^2\tag{4}\\
    \text{where} &\qquad y_i\text{: predicted value}\\
    &\qquad t_i\text{: actual/target value}\\
    &\qquad \text{Loss}_\text{MSE}\text{: summation of all training data}
\end{align*}\label{eq:loss_fun}

\begin{table}[h!]
    \centering
    \caption{CNN configuration for quantification of embedded crack length and position in HDPE.}
    \begin{tabular}{c|c|c|c|c|c}
        \hline
        Layer type & Channel & Kernel Size & Stride & Padding & Size / Neuron\\\hline
        Convolutional 1 & 4 & 8 & 4 & 2 & 440 x 4\\
        Convolutional 2 & 8 & 8 & 4 & 2 & 110 x 8\\
        Pooling 1 & 8 & 2 & 2 & 0 & 55 x 8\\
        FC 1 & - & - & - & - & 300\\
        FC 2 & - & - & - & - & 2\\
        \hline
    \end{tabular}
    
    \label{tab:cnn_config}
\end{table}

\subsection{CNN performance on simulations-generated testing signals}

\begin{figure}[h!]
    \centering
    \includegraphics[width=\textwidth]{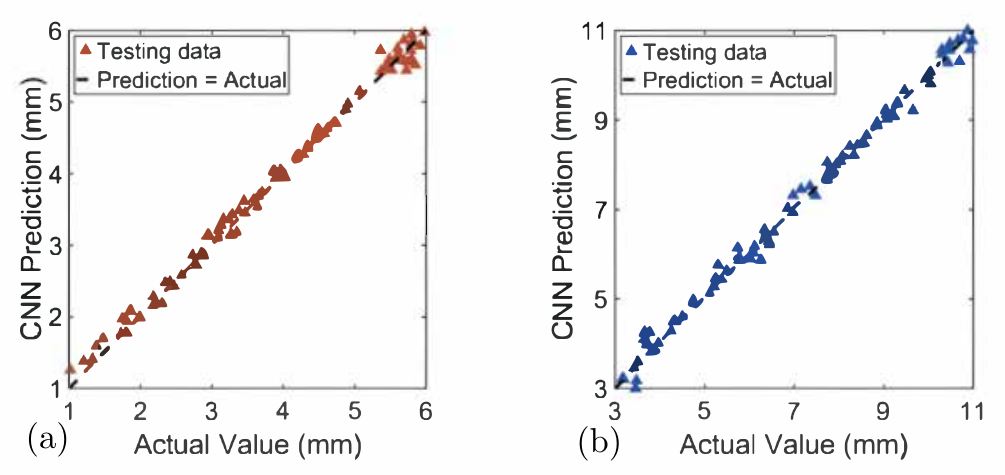}
    \caption{Performance of CNN on simulations generated testing data. Performance on crack length (left) and crack position (right). Points closer to the $45^\circ$ line indicate higher accuracy predictions.}
    \label{fig:testing_results}
\end{figure}

An additional 100 finite element simulated signals were generated as part of our simulation-based testing dataset. These signals were not passed to the CNN during training. Crack lengths and positions for this dataset were randomly chosen from the ranges introduced in Table \ref{tab:crack_range}.  Figure \ref{fig:testing_results} and Table \ref{tab:testing_errors} describe the predictive performance of trained CNN on the simulation-generated independent testing dataset. The x-axis in Figure \ref{fig:testing_results} denotes the actual values of the crack feature in the simulated geometry, and the y-axis denotes predicted values from the CNN. The dashed, black $45^\circ$ lines represent perfect predictions. The trained CNN demonstrates high accuracy for both crack length and position when provided independent finite element simulated testing data. The mean absolute percent errors (MAPE) of crack length and position are 3.2\% and 2.5\%, respectively, and the mean absolute errors (MAE) for length and position are 0.05 mm and 0.15 mm. These error values for the simulation testing data are small, indicating the CNN performed well.

\begin{table}[h!]
        \centering
        \caption{MAPE and MAE of CNN predictions on 100 simulated testing signals}
        \begin{tabular}{c|c|c}
            \hline
            {Parameter} & {MAPE (\%)} & {MAE (mm)}\\\hline
            {Length} & 3.2 & 0.05\\
            {Position} & 2.5 & 0.15\\
            \hline
        \end{tabular}
        \label{tab:testing_errors}
    \end{table}

\section{Validation Ultrasound Non-Destructive Experiments}\label{validation}

\subsection{Experimental setup}

Independent experimental validation of simulation-trained CNN is important for confirming the accuracy of our finite element simulations trained CNN method for predicting crack characteristics within real HDPE samples. We performed ultrasound NDT tests on fabricated HDPE specimens and applied previously simulation-trained CNN to predict the crack lengths and positions in HDPE test specimens. Test samples with embedded, non-visible cracks can be fabricated by 3D printing technology \cite{Niu2022SimulationMeasurements}. Compared to the conventionally processed bulk HDPE used in structural applications, 3D printed polymeric specimens would have significantly different attributes. HDPE experiences shrinking, voiding, and warping once extruded in this method \cite{Schirmeister20193DFabrication}. Additionally, the resulting 3D printed polymeric structure is highly porous. The high porosity of the structure can negatively affect ultrasound wave propagation and acoustic transmission. Since we are primarily focused on conventionally processed HDPE, we developed a test specimen fabrication method that did not rely on 3D printing.  

\begin{figure}[h!]
\centering
\includegraphics[width=\textwidth]{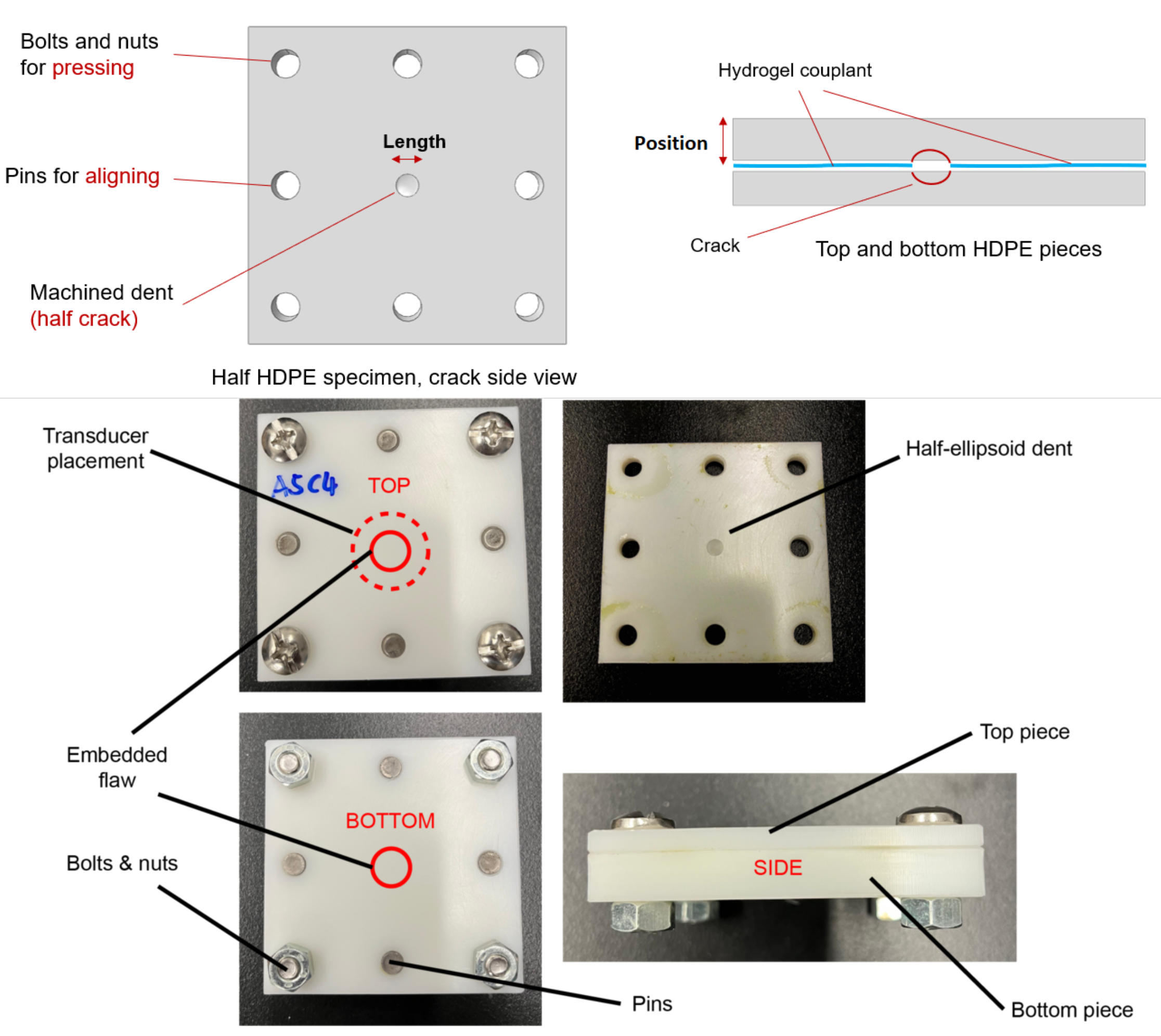}
\caption{The experimental design for HDPE specimen with an embedded crack (top) and one of the fabricated HDPE specimens for ultrasound testing following the design procedure (below).}
\label{fig:specimen}
\end{figure}

\begin{figure}[h!]
    \centering
    \includegraphics[width=0.85\textwidth]{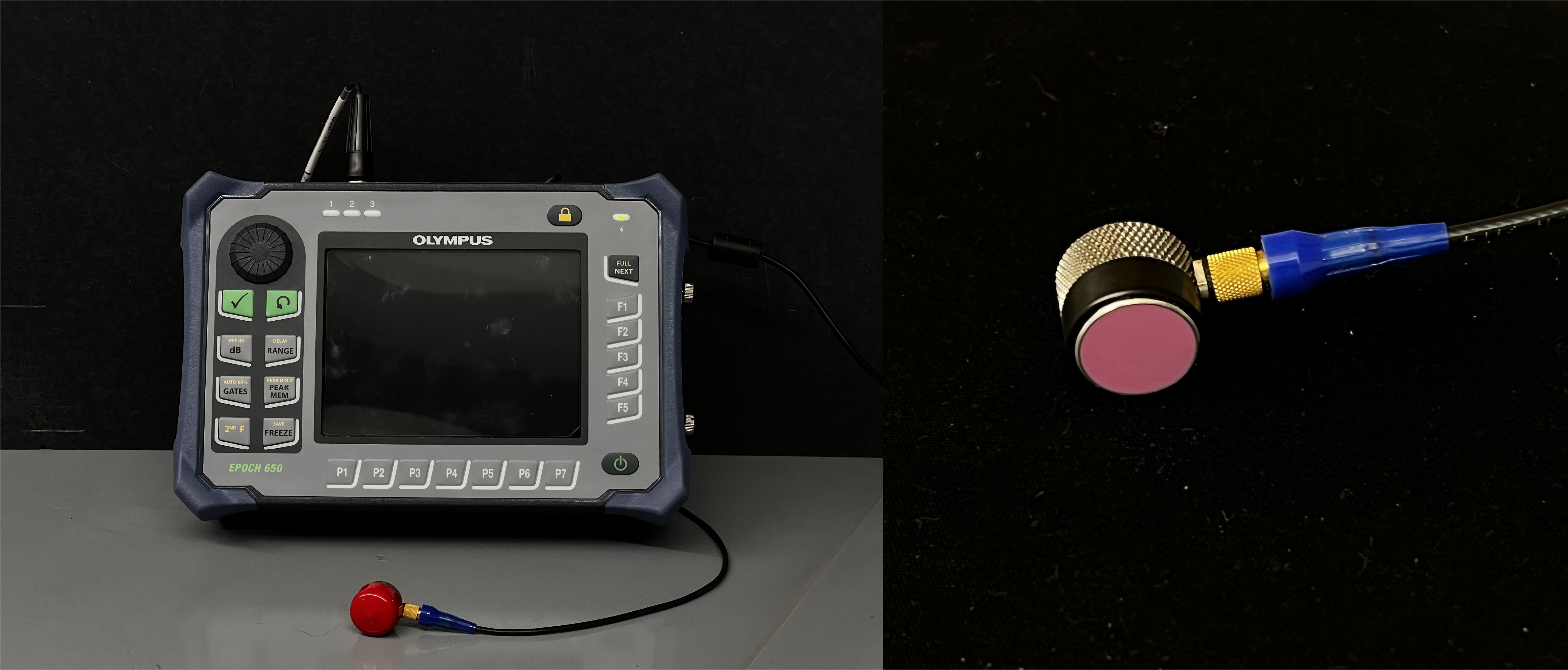}
    \caption{The Olympus Epoch 650 Ultrasound NDT Flaw Detector used in this study (left). The transducer has a 12.7 mm diameter and is a straight beam single element transducer (right).}
    \label{fig:ultrasound_equip}
\end{figure}

Our design for HDPE specimens with embedded cracks is illustrated in Figure \ref{fig:specimen}. Every test specimen was constructed using two flat sheets of HDPE, each with an in-plane length and width of 70 mm and varying thicknesses. The thickness of the two sheets were individually varied to change the through thickness crack position while keeping the overall specimen thickness at 12.7 mm. A half-ellipsoid dent was machined in the center of each sheet. Pressing the two sheets together with the dents facing each other results in an elliptical penny-shaped gap invisible from the exterior. This penny-shaped void serves to represent a non-visible, embedded crack. To ensure the two sheets were aligned once pressed together, eight 6.25 mm diameter holes were machined on the four edges and four corners. The corner holes served to hold tightly screwed bolts for the purpose of pressing and securing the two sheets together, while the edge holes held metal dowel pins to maintain alignment. The large $70\times70$ mm size and the dowel and bolt locations were chosen to keep waves reflected from the dowels, bolts and side boundaries outside the window of through thickness reflected wave measurements. These specimens and measurements represent through thickness measurements of embedded cracks in structures with large transverse spans (e.g., pipelines, storage tanks, etc.). 
To avoid potential air gaps in the mating surfaces, hydrogel couplant (35\% propylene glycol) was applied between the top and bottom sheets, then they were tightly pressed together\footnote{The sheets were separated to observe if any couplant was smeared into the crack. Any smeared couplant was removed from the crack indentation, and the process was repeated again until no gel was visible inside the crack.}.  The crack size and position were varied across fifteen fabricated HDPE specimens. The crack lengths selected were 2, 3, 4, 5, and 6 mm, and the crack positions were chosen to be 4, 7, and 10 mm from the surface of the specimen. Each specimen can be tested from both sides, resulting in thirty possible experiments. Five cases with very shallow crack positions (2.7 mm) were not considered because the crack echo and initial sensor pulse could overlap and interfere in these cases.
Removing these five experimental signals yields a final count of 25 independent experimental signals. All ultrasound experiments were conducted using an Olympus Epoch 650 Ultrasound NDT Flaw Detector, seen in Figure \ref{fig:ultrasound_equip}. The transducer used was a straight beam and single element with a frequency of 1 MHz and a sensor surface diameter of 12.7 mm. 

\subsection{Experimental results}

The finite element trained CNN was applied to twenty-five ultrasound signals obtained from NDT of the fabricated HDPE specimens with varying embedded crack lengths and positions. No experimental data was used to train the CNN, and these experiments were conducted solely to validate the proposed methodology for HDPE. 

\begin{figure}[h!]
\centering
\includegraphics[width=\textwidth]{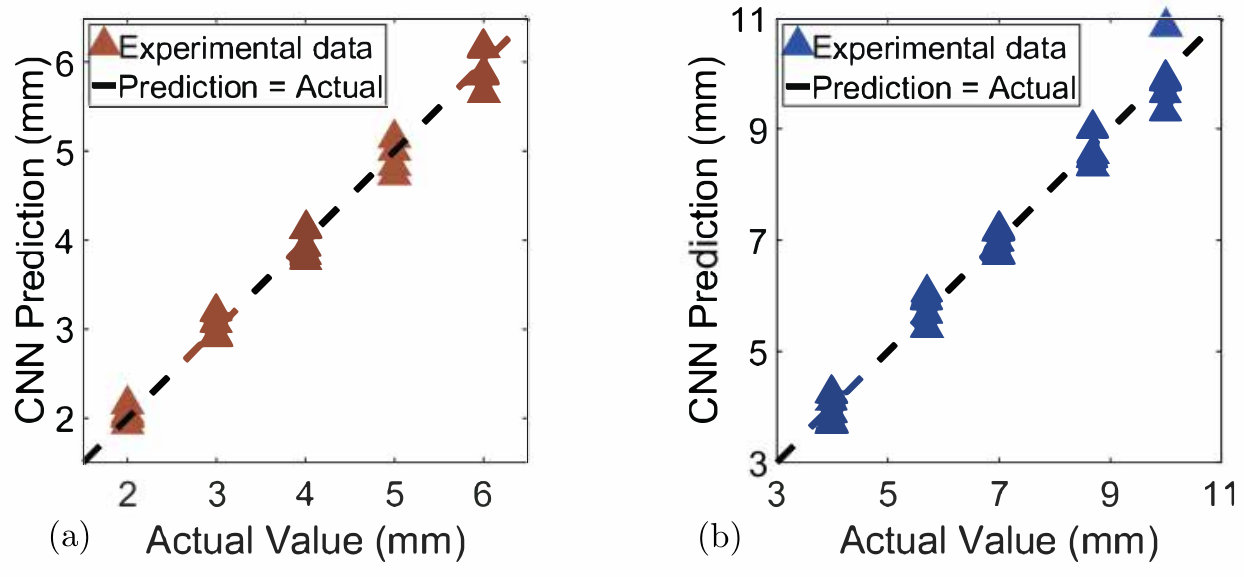}
\caption{CNN performance on ultrasound NDT experimental data. CNN predictive performance for (a) crack length and (b) crack position. Points closer to the $45^\circ$ line indicate higher accuracy predictions.}
\label{fig:specimen2}
\end{figure}

Figure \ref{fig:specimen2} presents the results of the CNN on the experimental signals. The $45\degree$ line represents a perfect fit. Points close to the line indicate the high accuracy of the predictions. These results demonstrate that the CNN is accurately predicting both crack length and crack position simultaneously from experimental ultrasound signals. Table \ref{tab:experi_full_result} lists the CNN predictions, true values, and percentage errors for the two crack characteristics. The mean absolute percent error (MAPE) was found to be 3.24\% for length and 3.78\% for position. The mean absolute error (MAE) for length and position was 0.07 and 0.26 mm, respectively. 

\begin{table}[h!]
    \centering
    \caption{Detailed summary of simulation-trained CNN performance on experimental ultrasound measurements on HDPE.}
    
\begin{tabular}{c|c|c|c|c|c|c}
\hline
\multirow{2}{*}{ Specimen
Number } & \multicolumn{3}{c}{Length (mm)} & \multicolumn{3}{c}{Position (mm)} \\
& Actual & Predicted & Error & Actual & Predicted & Error \\\hline
1 & 2 & 2.15 & 7.72\% & 4 & 3.69 & 7.82\% \\
2 & 2 & 2.05 & 2.52\% & 5.7 & 5.41 & 5.11\% \\
3 & 2 & 1.93 & 3.39\% & 7 & 6.79 & 2.95\% \\
4 & 2 & 2.02 & 0.98\% & 8.7 & 9.02 & 3.68\% \\
5 & 2 & 2 & 0.04\% & 10 & 9.65 & 3.48\% \\
6 & 3 & 3.07 & 2.46\% & 4 & 3.73 & 6.77\% \\
7 & 3 & 2.93 & 2.24\% & 5.7 & 5.66 & 0.73\% \\
8 & 3 & 2.94 & 1.93\% & 7 & 7.16 & 2.35\% \\
9 & 3 & 2.92 & 2.80\% & 8.7 & 8.33 & 4.20\% \\
10 & 3 & 3.19 & 6.41\% & 10 & 10.84 & 8.36\% \\
11 & 4 & 4.14 & 3.44\% & 4 & 3.88 & 3\% \\
12 & 4 & 3.84 & 3.93\% & 5.7 & 5.9 & 3.53\% \\
13 & 4 & 3.93 & 1.64\% & 7 & 6.74 & 3.71\% \\
14 & 4 & 3.78 & 5.46\% & 8.7 & 8.34 & 4.15\% \\
15 & 4 & 4.14 & 3.54\% & 10 & 9.86 & 1.35\% \\
16 & 5 & 4.83 & 3.34\% & 4 & 4.23 & 5.78\% \\
17 & 5 & 4.73 & 5.31\% & 5.7 & 6.05 & 6.12\% \\
18 & 5 & 4.73 & 5.40\% & 7 & 7.21 & 3.03\% \\
19 & 5 & 5.15 & 2.90\% & 8.7 & 8.56 & 1.57\% \\
20 & 5 & 5 & 0.07\% & 10 & 9.93 & 0.75\% \\
21 & 6 & 5.67 & 5.57\% & 4 & 4.1 & 2.45\% \\
22 & 6 & 5.89 & 1.83\% & 5.7 & 5.93 & 3.99\% \\
23 & 6 & 6.16 & 2.69\% & 7 & 6.98 & 0.27\% \\
24 & 6 & 6.18 & 2.97\% & 8.7 & 8.48 & 2.51\% \\
25 & 6 & 5.86 & 2.36\% & 10 & 9.32 & 6.75\% \\\hline
MAPE & \multicolumn{3}{c}{3.24\%} & \multicolumn{3}{c}{3.78\%} \\
MAE & \multicolumn{3}{c}{0.07 mm} & \multicolumn{3}{c}{0.26 mm} \\
\hline
\end{tabular}
    \label{tab:experi_full_result}
\end{table}

\section{Conclusions}\label{conclusions}

\begin{figure}[h!]
\centering
\includegraphics[width=0.8\textwidth]{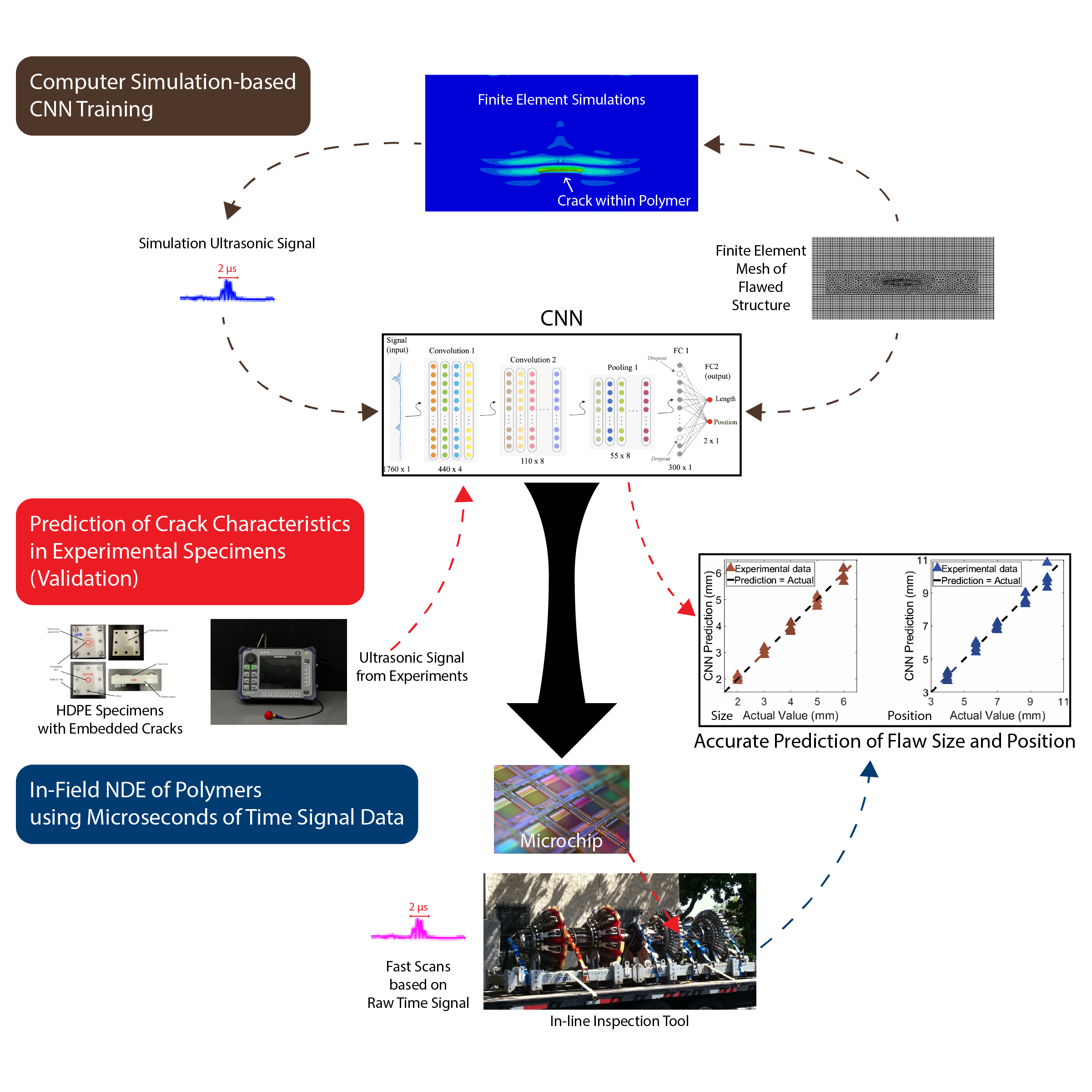}
\caption{NDE method for crack measurements from microseconds of ultrasound time signal and its potential application in structures like long pipelines using In-Line-Inspection (ILI) devices. Images with permission from \cite{Web_Ockel,Web_Lemonn}}
\label{fig:overview}
\end{figure}

Flaws can manifest in polymers during fabrication or operations. As the demand for polymers increases, so does the need for rapid and accurate characterization of crack length and position to avoid sudden catastrophic failures. Processing and interpreting ultrasound signals using machine learning techniques have demonstrated significant potential, especially for detecting non-visible, embedded flaws. Machine learning has been applied to image based ultrasound applications. Our finite element trained CNN method for HDPE is based on ultrasound time amplitude signal and not based on image analysis. This is an important distinction as the 2D image rendering NDT process is very slow and is limited as it is susceptible to losing subtle crack features information during the post-processing required to create images. The ultrasound time signal data with measurement windows of tens of microseconds at each location is convenient for measuring in large structures. Our study confirms that acoustic attenuation and dispersion due to viscoelasticity in HDPE can be reasonably neglected for ~1 MHz frequency and relatively shorter signal travel distances ($\sim$50 mm).  We conducted 3D finite element simulations representing ultrasonic A-scan wave propagation inside HDPE with embedded cracks to show that simulation-generated ultrasound signals can train signal-based CNN well. We demonstrate that the simulation-trained CNN can subsequently predict the crack length and position of penny-shaped embedded cracks in real HDPE samples with very good accuracy. In our study, the average error in crack length and position predictions was less than 3.8\%. As summarized in Figure \ref{fig:overview}, this study suggests a method for HDPE that can potentially be extended to other solid polymers where a CNN is trained using geometry and flaw type specific finite element ultrasound wave propagation simulations and then applied onboard an inspection tool to detect and characterize the specific flaws that it is trained for. 

\section*{Declaration of Competing Interest} The authors declare no conflict of interest.

\section*{Data Availability} The data will be available upon request. 

\section*{Acknowledgement}
We gratefully acknowledge support from the U.S. Department of Transportation to V. Srivastava under grant number 693JK32050001CAAP. The views and opinions expressed in this article are those of the authors and do not necessarily reflect the official policy or position of any agency of the U.S. government.


\renewcommand\refname{ References}
\bibliographystyle{elsarticle-num.bst}
\bibliography{hdpe_references} 
\end{document}